\documentclass{article}

\usepackage{arxiv}

\usepackage[utf8]{inputenc} 
\usepackage[T1]{fontenc}    
\usepackage{hyperref}       
\usepackage{url}            
\usepackage{booktabs}       
\usepackage{amsfonts}       
\usepackage{nicefrac}       
\usepackage{microtype}      
\usepackage{lipsum}   
\usepackage{graphicx}
\usepackage{doi}
\usepackage{amsmath,amsfonts,amssymb}
\usepackage{array,multirow,graphicx}
\usepackage{comment}

\title{Sharing Generative Models Instead of Private Data: \\  A Simulation Study on Mammography Patch Classification

}

\author{
  Zuzanna Szafranowska\thanks{ equal contribution. } \\ 
  Faculty of Mathematics and Computer Science \\ 
  University of Barcelona \\ 
  Spain \\ 
  \texttt{z.szafranowska@gmail.com} \\
   \And
  {Richard Osuala$^*$}  \\ 
  Faculty of Mathematics and Computer Science \\ 
  University of Barcelona \\ 
  Spain \\ 
  \texttt{richard.osuala@gmail.com} \\
   \And
  Bennet Breier \\
  Faculty of Mathematics and Computer Science \\ 
  University of Barcelona \\ 
  Spain \\ 
   \And
  Kaisar Kushibar\\
  Faculty of Mathematics and Computer Science \\ 
  University of Barcelona \\ 
  Spain \\ 
   \And
    Karim Lekadir \\
  Faculty of Mathematics and Computer Science \\ 
  University of Barcelona \\ 
  Spain \\ 
   \And
  Oliver Diaz \\
  Faculty of Mathematics and Computer Science \\ 
  University of Barcelona \\ 
  Spain \\ 
}

\date{}

\renewcommand{\headeright}{accepted at IWBI 2022}
\renewcommand{\undertitle}{PREPRINT}
\renewcommand{\shorttitle}{\textit{Sharing Generative Models Instead of Private Data}}

\hypersetup{
pdftitle={Sharing Generative Models Instead of Private Data: A Simulation Study on Mammography Patch Classification},
pdfauthor={Zuzanna Szafranowska, Richard Osuala, Bennet Breier, Kaisar Kushibar, Karim Lekadir, Oliver Diaz},
pdfkeywords={Generative Adversarial Networks, Transformer, Deep Learning, Synthetic Data, Patient Privacy, Mammography, Breast Cancer},
}

\begin{document} 

\maketitle

\newcommand\blfootnote[1]{%
  \begingroup
  \renewcommand\thefootnote{}\footnote{#1}%
  \addtocounter{footnote}{-1}%
  \endgroup
}

\begin{abstract}
Early detection of breast cancer in mammography screening via deep-learning based computer-aided detection systems shows promising potential in improving the curability and mortality rates of breast cancer. However, many clinical centres are restricted in the amount and heterogeneity of available data to train such models to (i) achieve promising performance and to (ii) generalise well across acquisition protocols and domains. As sharing data between centres is restricted due to patient privacy concerns, we propose a potential solution: sharing trained generative models between centres as substitute for real patient data. In this work, we use three well known mammography datasets to simulate three different centres, where one centre receives the trained generator of Generative Adversarial Networks (GANs) from the two remaining centres in order to augment the size and heterogeneity of its training dataset. We evaluate the utility of this approach on mammography patch classification on the test set of the GAN-receiving centre using two different classification models, (a) a convolutional neural network and (b) a transformer neural network. Our experiments demonstrate that shared GANs notably increase the performance of both transformer and convolutional classification models and highlight this approach as a viable alternative to inter-centre data sharing. \blfootnote{This is a preprint of a paper accepted as oral presentation at the International Workshop on Breast Imaging (IWBI) 2022}

\end{abstract}


\section{INTRODUCTION}
\label{sec:intro}  

With an estimated worldwide incidence rate of 47.8 per 100,000 people in 2020 (both sexes, all ages), breast cancer is the cancer type with the highest prevalence in the world. It accounts for an estimated 2.22 million new cases and 684,996 deaths each year \cite{globalCancerObservatory}.

Screening mammography (MMG) provides early detection and contributes to reducing breast cancer mortality\cite{Abdelrahman2021}.

However, MMG images are limited by their error rates due to tissue superposition which could lead to underdiagnosis of significant breast cancers (false negatives) and overdiagnosis of insignificantly abnormal or healthy cases (false positives) \cite{Abdelrahman2021, Osuala2021}.
In this regard, deep learning based computer-aided detection (CADe) systems have shown great promise in improving and automating the decision making process of mammograms \cite{becker2017deep, Abdelrahman2021}.
However, deep learning methods are known to require large amounts of training data to achieve accurate, reliable and robust performance.

Scarcity of expert-annotated medical images often constrains deep learning based methods to be trained and evaluated on a small dataset coming from a single centre~\cite{castro2020causality}. Accordingly, such methods suffer from lack of generalisation and robustness~\cite{lekadir2021future}. A solution to this problem is inter-centre data sharing. However, clinical centres are constrained from sharing sensitive patient data due to technical, legal, and most importantly, ethical concerns\cite{bi2019artificial, Osuala2021}.

Synthetic images generated by Generative Adversarial Networks (GANs) \cite{goodfellow2014generative} have been shown to be of high perceived visual realism for mammography 
and to improve downstream tasks including cancer detection
, tumour segmentation and classification 
\cite{Osuala2021}. Therefore, in this work, we hypothesise that GANs can overcome the inter-centre data sharing constraints. After learning the real data distribution, GANs can generate synthetic data with limited risk of revealing sensitive patient information \cite{shin2018medical, Osuala2021}. Then, a clinical centre can share a trained generative model that will act as a proxy for the real patient data.

\begin{figure} [ht]
\begin{center}
\includegraphics[height=5cm]{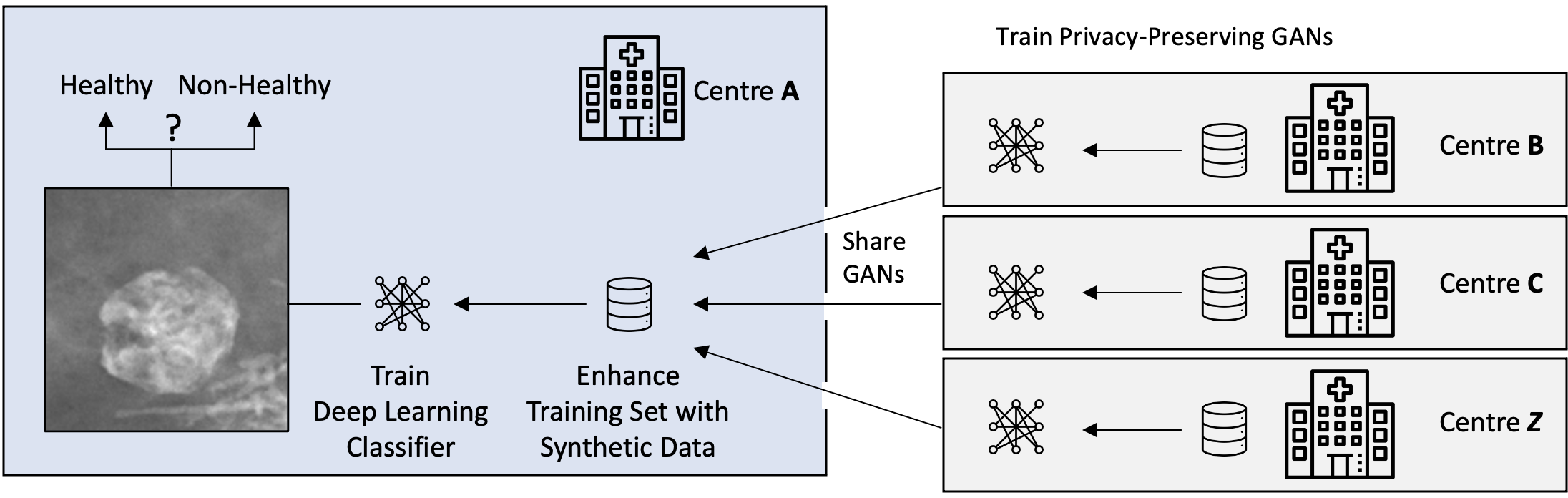}
\end{center}
\caption[setup] 
   { \label{fig:setup} 
Overview of sharing generative models (e.g., GANs) as substitutes for multi-centre patient data to (i) overcome data-scarcity in single centres, and (ii) to enhance prediction performance and robustness. The latter is exemplified on the task of classifying the presence of a lesion in a mammography patch.}
\end{figure}

In this work, we investigate the application of GANs as substitutes for multi-centre real patient data. Using three well known mammography datasets acquired at different sites from Portugal \cite{Moreira2012, Lopez2012} and UK \cite{halling2020optimam}, we simulate three different centres, where one centre receives GANs from the two remaining centres to augment the size and heterogeneity of its training dataset. 
Figure~\ref{fig:setup} shows our simulated privacy-preserving data sharing setup using GANs. We evaluate the performance of a convolutional neural network (CNN) \cite{Fukushima1980, lecun1998gradient} and a transformer neural network \cite{vaswani2017attention} for healthy versus non-healthy tissue classification in a region of interest (ROI) using two training strategies: 1) using data from only a single centre; and 2) using additional synthetic data from other centres. We demonstrate through our experiments that augmenting single centre data using GAN generated image ROIs considerably improves classification performance.

\section{METHODOLOGY}
\label{sec:methodology}

Our methodological framework consists of two parts. First, we utilise a GANs to learn the distribution of the data from data-providing centres (Centres B, C, etc. in Figure~\ref{fig:setup}). These trained GANs from each centre are then sent to the main centre (Centre A in Figure~\ref{fig:setup}) for further use. Second, the main centre trains a classification model that classifies whether a lesion is present in MMG patches. 

The training data for the classification model in the main centre is augmented by GAN-generated lesion patches trained on data from other centres. To maintain class balance between \textit{healthy} and \textit{non-healthy} (i.e., has lesion) labels, we add the same number of healthy patches extracted from existing healthy control MMGs from centre A.
Finally, the performance of the classification model trained only on data from centre A is compared to the performance of the same classification model trained on the combination of centre A data and synthetic data from other centres.

In our experiments, we use INbreast\cite{Moreira2012} and BCDR\cite{Lopez2012} mammography datasets and an additional pretrained GAN trained on the OPTIMAM dataset~\cite{halling2020optimam}. INbreast consists of digital MMGs from 115 patients, totalling 410 images. A total number of lesions is 3028 (including 116 masses), which corresponds to the number of INbreast patches in our experiments that include both benign and malignant masses and calcifications. The BCDR dataset consists of both digital and film-scanned MMG images from 1010 patients, totalling 1493 lesions (including 639 masses). The pretrained GAN was trained on 2215 masses extracted from the OPTIMAM dataset. 
 
In particular, we gathered experimental results for testing on data from centre A (represented by the INbreast dataset) after training on data from (i) centre A (INbreast), (ii) centre A (INbreast) and centre B (BCDR), (iii) centre A (INbreast) and synthetic data from centre B (BCDR) and/or centre C (OPTIMAM).
We conducted these experiments (i to iii) for two scenarios. First, the scenario where 100\% of the centre A (INbreast) training data is used for training. And secondly, the setting where centre A is experiencing data scarcity with its internal training data set randomly reduced by 50\%. Also we separately conduct each experiment for tumour masses specifically, and for breast lesions (calcifications, masses, etc.) in general. 
As shared generative model of centre B, we train both a Deep Convolutional Generative Adversarial Network (DCGAN) \cite{radford2015unsupervised} and a Wasserstein GAN with Gradient Penalty (WGAN-GP)\cite{gulrajani2017improved}, while for centre C we reuse a pretrained DCGAN from Alyafi et al \cite{alyafi2020dcgans, alyafi20quality} via the \textit{medigan} model sharing library\cite{richard_osuala_2022_6327625}\footnote{\href{https://medigan.readthedocs.io/}{https://medigan.readthedocs.io/}}. 
We train, evaluate, and report results for two different classification models in centre A, namely (a) a convolutional neural network (CNN) and (b) a Swin transformer neural network \cite{liu2021swin}. 

The classification performance is evaluated using the accuracy, F1-score, area under receiver operating characteristic curve (AUROC), and area under precision-recall curve (AUPRC) metrics. To further increase the informative value of our results, we run all experiments three times with a different random seed in each run and report the resulting mean and standard deviation of each metric.

All classification experiments were run on a machine equipped with NVIDIA GTX 1070 8GB GPU, using the PyTorch library~\cite{pytorchPaszke}. Our GANs were trained on a NVIDIA RTX 2080 Super 8GB GPU, also using PyTorch.

\subsection{Mammogram Patch Extraction}
Healthy patches are extracted from INbreast MMGs of healthy breasts that have no annotation that indicates the presence of a lesion. We generate a number of bounding boxes randomly defined within an entirely healthy breast image, ensuring that these patches never contain more than 40\% of background pixels.

Non-healthy patches are crops containing both malignant and benign lesions of any breast imaging-reporting and data system (BI-RADS) score and of any lesion type present in the datasets including masses, calcifications, microcalcifications, and architectural distortions. 
We use the lesions contour information specified in the original datasets~\cite{Lopez2012, Moreira2012} and create a bounding box enclosing it. Then, we create a square patch around that bounding box, ensuring a margin of 60 pixel in each direction from the lesion bounding box. If the margin extends beyond the mammogram's border, a translation is performed to the patch to ensure it is fully within the mammogram's limits.

After specifying the patch bounding boxes, we use the same pre-processing routine at training time for both healthy and non-healthy patches to ensure that class-specific pre-processing artefacts are not introduced that would otherwise be easily distinguishable by the classifier.
Each patch, defined by its bounding boxes, is first zoomed and then translated by normally random amounts. In doing so, we offset the patch from its original bounding box, which would otherwise always be close to the centre of the patch, and as a strategy of data augmentation. 
Finally, each patch is resized to 128x128 pixels using inter-area interpolation, whereby image ratios are maintained.

\subsection{Generative Adversarial Networks as Data Substitute}
GANs \cite{goodfellow2014generative} are a type of generative model and are composed of the discriminator (D) and the generator network (G) that compete against each other in a two-player zero-sum game defined by the value function shown in equation \ref{eq1}. 
\\
\begin{equation}\label{eq1}
\begin{aligned}
\min_{G} \max_{D} V(D,G) = \min_{G} \max_{D}[\mathbb{E}_{x\sim \mathbb{P}_{data}} [log (D(x))] + \mathbb{E}_{z\sim \mathbb{P}_{z}} [log(1 - D(G(z)))]]
\end{aligned}
\end{equation}

\begin{figure} [ht]
\begin{center}
\includegraphics[height=8cm]{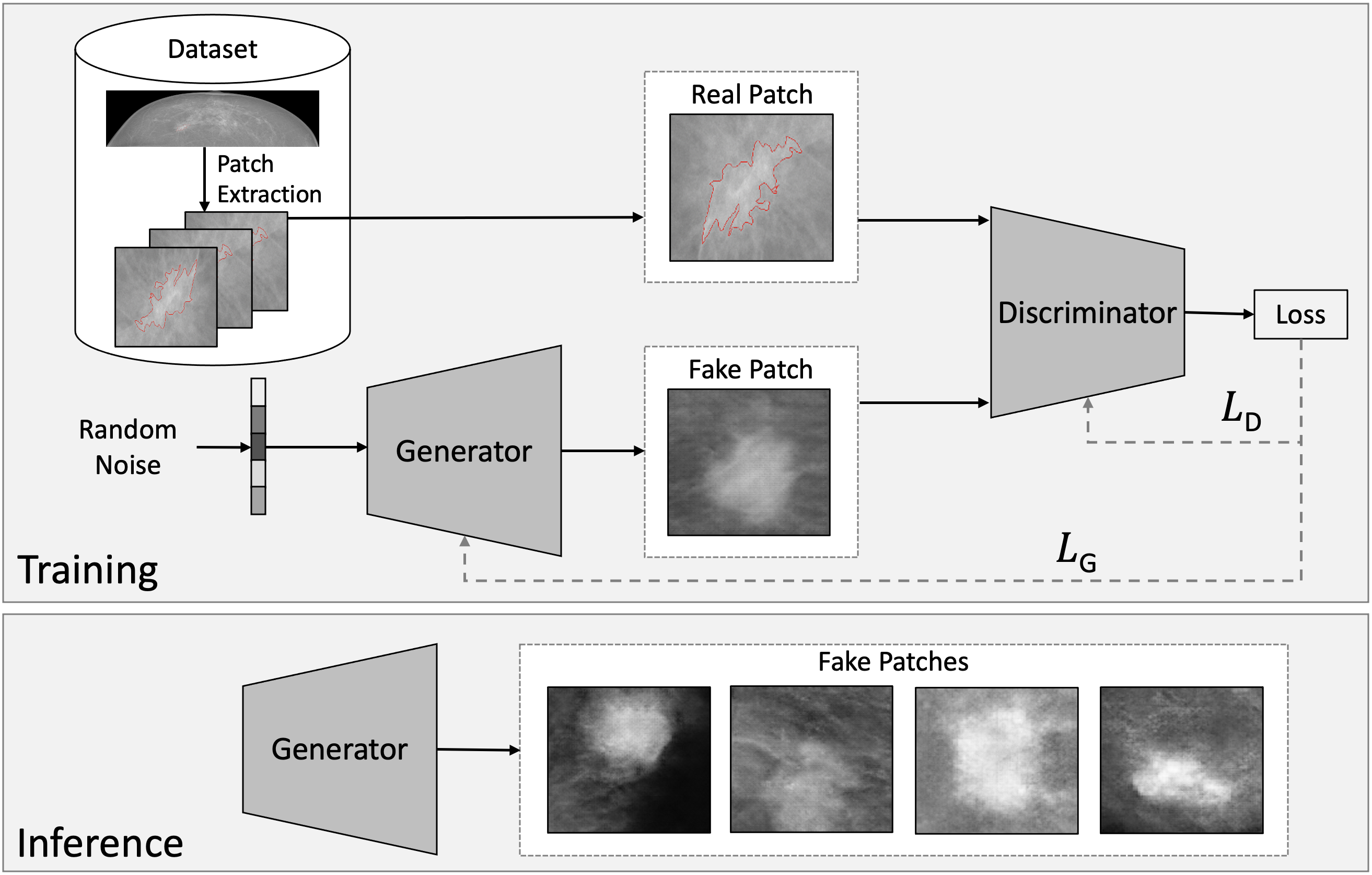}
\end{center}
\caption[gans] 
   { \label{fig:gans} 
GAN training: After patches of regions-of-interest (e.g., containing lesions, masses) are extracted from the mammography dataset, the discriminator strives to distinguish real from generated patches.
GAN inference: The generator is able to produce new samples from the training data distribution.
}
\end{figure}

After GAN training, the generator model is extracted and is able to produce samples from the training distribution, as depicted by figure \ref{fig:gans}.

\subsubsection{DCGAN Generator Trained on BCDR Data} \label{dcgan_bcdr}
We adopt a DCGAN~\cite{radford2015unsupervised} with some adjustments suggested in \cite{alyafi2020dcgans} such as one-sided label smoothing in range [0.8, 1.1], and a discriminator with a kernel size 6 instead of 4. We train our DCGAN to learn the distribution of the training data consisting of 128x128 pixel grayscale mammogram patches. The DCGAN learns a mapping between a vector containing 100 numerical values to a mammogram patch containing a breast lesion. The GAN training data is augmented by random (p=0.5) horizontal and (p=0.5) vertical flipping and uses a batch size of 16. Depending on the classification objective of our experiments the GAN is either trained on patches of all lesion types, or specifically on patches containing a mass.

We train our DCGAN on the BCDR dataset for 3000 epochs to generate synthetic patches similar to the real patches. After each epoch, we visually assessed the fidelity of the mammogram patches generated from a set of fixed noise vectors. During training we noticed an increase in fidelity, but also a decrease in diversity of the generated patches: The similarity of some of the generated lesions indicated the occurrence of mode collapse, the state where the generator has learned to repeatedly generate a limited subset of samples to fool the discriminator. Interestingly, the same fixed noise vector input created varying lesion shapes and textures in different training iterations indicating a high diversity across iterations and epochs.

Observing this behaviour two measures were taken. 
Firstly, to maximise diversity, we store the weights of our DCGAN during training on each 50th epoch starting in epoch 500. After training, we generate mammogram patches using the stored weights (epoch 500 to 3000) to generate the synthetic dataset. 
Secondly, we explore further GAN alternatives less prone to mode collapse and, based on our DCGAN network architecture, implement Wasserstein GAN with Gradient Penalty (WGAN-GP)~\cite{gulrajani2017improved}. 

\subsubsection{WGAN-GP Generator Trained on BCDR Data}
\label{wgangp_bcdr}
To overcome mode collapse in DCGAN and to increase training stability, we follow the approach of \cite{magister2021generative} of substituting DCGAN's binary cross-entropy loss with a Wasserstein distance based loss function.
We apply the Wasserstein with gradient penalty~\cite{gulrajani2017improved} loss function to the setup described in \ref{dcgan_bcdr}. Equation \ref{eq2} displays the WGAN-GP loss function with penalty coefficient $\lambda$ (set to 10 in our experiments) and distribution $\mathbb{P}_{\hat x}$ sampling uniformly along straight lines between
pairs of points from the generator distribution $\mathbb{P}_{g}$, and the data distribution $\mathbb{P}_{data}$.
\begin{eqnarray}\label{eq2}
\begin{array}{l}
 L \ = \ {E_{\tilde x \sim {\mathbb{P}_g}}}\left[ {D\left( {\tilde x} \right)} \right] \ - \ {E_{x \sim {\mathbb{P}_{data}}}}\left[ {D\left( x \right)} \right] \ + \ \lambda \ {E_{\hat x \sim {\mathbb{P}_{\hat x}}}}\left[ {{{\left( {{{\left\| {{\nabla _{\hat x}}D\left( {\hat x} \right)} \right\|}_2} - 1} \right)}^2}} \right] \\ 
 \end{array}
 \end{eqnarray}
The discriminator (alias critic) is updated 5 times per generator update. Further, we remove the batch normalization layers from the DCGAN discriminator, as WGAN-GP penalises the norm of the discriminator's gradient per input sample rather than per batch. Both gradient penalty (WGAN-GP) ~\cite{gulrajani2017improved} and weight clipping (WGAN) \cite{arjovsky2017wasserstein} enforce a 1-Lipschitz constraint while the former additionally avoids model capacity underuse, which motivates our choice of WGAN-GP. We train WGAN-GP for 10000 epochs on mammogram patches that contain masses and for 2700 epochs on mammogram patches that contain any type of lesion. In both cases, due to high image diversity upon visual assessment, only the checkpoint from the last epoch was used to generate the synthetic images for our subsequent classification experiments.

\subsubsection{DCGAN Generator Pretrained on OPTIMAM Data}
\begin{figure} [ht]
\begin{center}
\includegraphics[height=5.2cm]{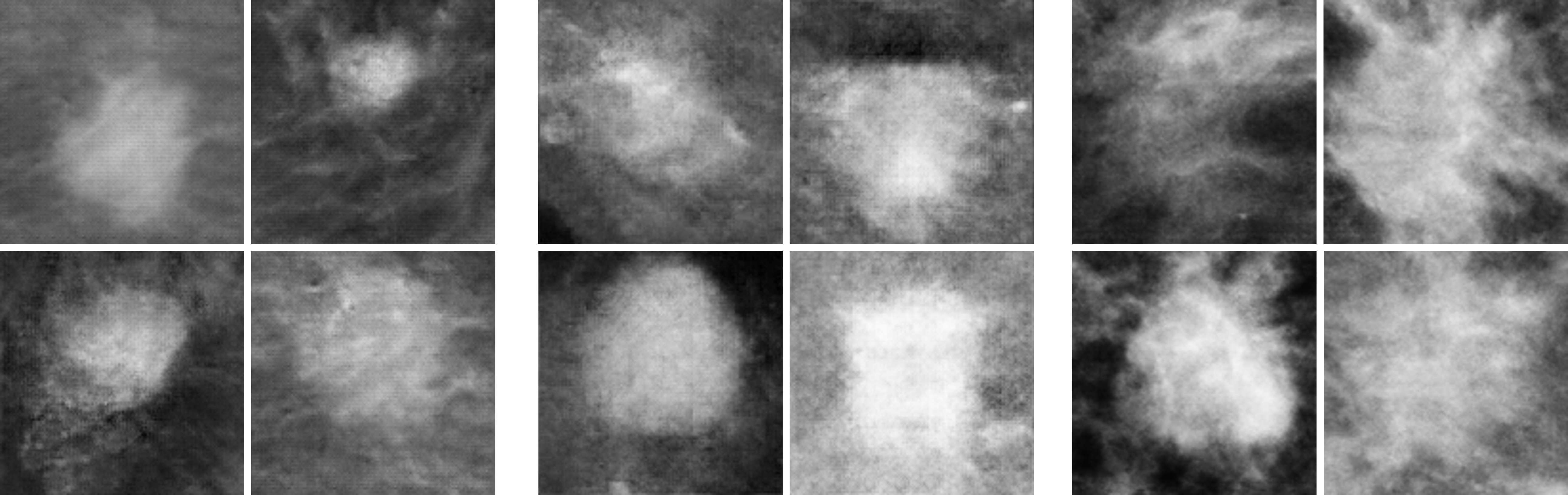}
\end{center}
\caption[synth] 
   { \label{fig:synth} 
Left column: synthetic patches generated from BCDR (using our DCGAN).
Middle column: synthetic patches generated from BCDR (using our WGAN-GP).
Right column: synthetic patches generated from OPTIMAM (using the pretrained DCGAN of Alyafi et al~\cite{alyafi2020dcgans, alyafi20quality}).}
\end{figure}

Furthermore, we use the DCGAN published by Alyafi et al~\cite{alyafi2020dcgans, alyafi20quality} from the \textit{medigan} \cite{richard_osuala_2022_6327625} library pre-trained on 2215 mass patches from the OPTIMAM dataset to generate an additional dataset of synthetic mammogram patches containing masses. Figure \ref{fig:synth} illustrates manually selected synthetic images generated by Alyafi et al's DCGAN, and our BCDR-trained DCGAN (\ref{dcgan_bcdr}) and WGAN-GP (\ref{wgangp_bcdr}). As opposed to our GANs, Alyafi et al's DCGAN only generates patches containing a mass and is not trained on other types of lesions. 

\subsection{Classification Models}

All experiments with the classifiers are validated and tested on INbreast~\cite{Moreira2012} data that have been split previously into training (approx. 70\%), validation (approx. 15\%) and testing (approx. 15\%) subsets. We ensure that (1) all images from a single patient are always in only one of the subsets, (2) healthy and non-healthy patches are always present in similar amounts, and (3) the distributions of breast density values in each of the subsets are as similar as possible. Dataset splits are performed separately, once for the experiments where we classify any type of lesion (masses, calcifications, etc) and once for the experiments where we classify exclusively the lesion type ``mass''.

In the cases when the patches with lesions stem from two GANs, we add half the patches from each GAN. Additionally, for all types of augmentations, we add as many healthy patches from INbreast as needed to maintain the class balance between healthy and non-healthy training data. Overall, we extract 1374 (70 in mass classification) real patches for each class from INbreast for training. Then, 1200 additional synthetic patches, either all lesions or only masses and either from one or from two centres are added to the training as augmentation. 
 
In cases of augmentation with real BCDR data we extract patches from both digital and film-scanned BCDR mammograms and add 1148 patches for lesion classification and 500 patches for mass classification to the class ``non-healthy''.

\subsubsection{CNN Classification Model}
As a baseline classification model, we implement a CNN, consisting of four convolutional followed by two fully-connected layers, with ReLU~\cite{reluAgarap2018deep} activation functions.
After each convolutional layer and after the first fully-connected layer, batch normalization~\cite{ioffe2015batch} is performed. Also, dropout~\cite{dropout} is applied between the two fully-connected layers to introduce regularization during training and reduce the risk of overfitting. Finally, the network's output is passed through a softmax function followed by a logarithm. Our loss function is defined as cross-entropy.
For training the network, we first initialise the weights randomly after setting a random seed. Then, we use stochastic gradient descent optimization of the loss function with the learning rate of 0.001 and momentum of 0.9. 
To obtain the best model, after each epoch, we perform validation on a separate subset of data, and only save the model if it achieves superior AUPRC score as compared to the saved models from previous epochs. The upper limit of epochs is set to 100. 

\subsubsection{Swin Transformer Classification Model}

For further corroboration of our results, we run the same experiments using Swin Transformer~\cite{liu2021swin} as classification model. This increases the generalisability of our approach by comparing two considerably different methodological frameworks that showed state-of-the-art performances in vision tasks. Due to heavy exploitation of self-attention mechanisms, Transformers consider relations between all pair-wise local regions in the image. Accordingly, they do not assume that related features are close to each other in the image and eliminate such inductive bias that is present in CNNs. Swin Transformers are an extension of hierarchical vision transformers, which implements shifted windows mechanism as a way to limit self-attention computation to non-overlapping local windows, while allowing for cross-window connection. In our experiments, we used the original Swin Transformer setup as in Liu et al~\cite{liu2021swin}. Therefore, we resized all the input patches by resampling with respect to pixel area relation to $224\times 224$ and stacked them to obtain a three channel input.

\section{RESULTS}
\label{sec:results}

Tables~\ref{tab:results1}~and~\ref{tab:results2} summarise our experimental results. Each experiment yields insights into different combinations of datasets, along three dimensions:
First, we use different data augmentations for the training set, namely synthetic patches only from BCDR generated either with a WGAN-GP or a DCGAN, synthetic patches only from OPTIMAM (DCGAN), synthetic patches from both BCDR and OPTIMAM where for BCDR we used either a DCGAN or a WGAN-GP, or real patches from BCDR.
Including real patches from BCDR serves as an expected upper bound for the experiments with synthetic patches.
Second, we train the classifier once with the entire INbreast training set and once with only 50\% thereof.
Third, we either include all lesion-types in the non-healthy class or masses only.

\begin{table}[ht]
\caption{Results for classification of \textbf{all lesion types} including masses, calcifications, etc, using CNN (top) and Swin Transformer (bottom). The left-most column refers to the source of data added to the INbreast training set.
``Both (a)'' means augmenting with synthetic patches from both BCDR (WGAN-GP) and OPTIMAM (DCGAN), while ``Both (b)'' means BCDR (DCGAN) and OPTIMAM (DCGAN). 
All classifiers are tested on the INbreast testset.
The best results per column are presented in bold font. 
Results are shown as mean(std).} 

\label{tab:results1}
\begin{center}
\scalebox{0.70}{
\begin{tabular}{|c|l||c|c|c|c||c|c|c|c|} 
\hline
\rule[-1ex]{0pt}{3.5ex} & & \multicolumn{4}{c||}{100\% of the INbreast training data} & \multicolumn{4}{c|}{50\% of the INbreast training data} \\ \cline{2-10}
\rule[-1ex]{0pt}{3.5ex} & Augmentation & Accuracy & F1 & AUROC & AUPRC & Accuracy & F1 & AUROC & AUPRC  \\ \hline
\hline
\multirow{5}{*}{\parbox[t][12mm][b]{2.5mm}{\rotatebox[origin=c]{90}{CNN}}}
\rule[-1ex]{0pt}{3.5ex} & None & 0.942(.005) & 0.693(.020) & 0.955(.007) & 0.880(.014) & 0.933(.010) & 0.661(.035) & 0.960(.004) & 0.878(.013) \\ \cline{2-10}
\rule[-1ex]{0pt}{3.5ex} & BCDR (WGAN-GP) & 0.938(.005) & 0.673(.012) & 0.957(.004) & 0.839(.047) & 0.929(.048) & 0.669(.130) & 0.953(.003) & 0.863(.015) \\ \cline{2-10}
\rule[-1ex]{0pt}{3.5ex} & BCDR (DCGAN) & 0.936(.030) & 0.685(.099) & \textbf{0.966(.001)} & \textbf{0.887(.006)} & 0.926(.034) & 0.652(.110) & 0.960(.005) & 0.874(.006) \\ \cline{2-10}
\rule[-1ex]{0pt}{3.5ex} & OPTIMAM (DCGAN) & \textbf{0.945(.005)} & 0.701(.014) & 0.961(.007) & 0.871(.010) & 0.945(.002) & 0.703(.004) & 0.963(.003) & 0.879(.007) \\ \cline{2-10}
\rule[-1ex]{0pt}{3.5ex} & Both (a) & 0.943(.010) & 0.701(.039) & 0.965(.004) & 0.863(.014) & \textbf{0.947(.020)} & \textbf{0.718(.075)} & \textbf{0.965(.005)} & \textbf{0.886(.014)} \\ \cline{2-10}
\rule[-1ex]{0pt}{3.5ex} & Both (b) & \textbf{0.945(.017)} & \textbf{0.704(.065)} & 0.959(.005) & 0.871(.013) & 0.945(.013) & 0.706(.044) & 0.964(.007) & \textbf{0.886(.013)} \\ \cline{2-10}
\rule[-1ex]{0pt}{3.5ex} & Real BCDR & 0.924(.001) & 0.633(.001) & 0.958(.001) & 0.851(.003) & 0.938(.007) & 0.680(.028) & \textbf{0.965(.002)} & 0.884(.016) \\
\hline
\hline
\multirow{5}{*}{\parbox[t][16mm][b]{2.5mm}{\rotatebox[origin=c]{90}{Swin Transformer}}}
\rule[-1ex]{0pt}{3.5ex} & None & 0.914(.065) & 0.634(.161) & 0.951(.004) & 0.859(.010) & 0.949(.040) & 0.734(.138) & 0.933(.004) & 0.860(.006) \\ \cline{2-10}
\rule[-1ex]{0pt}{3.5ex} & BCDR (WGAN-GP) & 0.912(.082) & 0.647(.186) & 0.952(.001) & 0.864(.002) & \textbf{0.974(.005)} & 0.824(.027) & 0.937(.015) & 0.860(.006) \\ \cline{2-10}
\rule[-1ex]{0pt}{3.5ex} & BCDR (DCGAN) & \textbf{0.974(.002)} & \textbf{0.826(.009)} & 0.955(.004) & 0.874(.005) & 0.959(.030) & \textbf{0.880(.066)} & \textbf{0.958(.007)} & \textbf{0.916(.067)} \\ \cline{2-10}
\rule[-1ex]{0pt}{3.5ex} & OPTIMAM (DCGAN) & 0.946(.038) & 0.724(.120) & 0.957(.002) & 0.870(.002) & 0.927(.043) & 0.667(.150) & 0.953(.016) & 0.871(.029) \\ \cline{2-10}
\rule[-1ex]{0pt}{3.5ex} & Both (a) & 0.922(.040) & 0.637(.112) & 0.940(.019) & 0.852(.022) & 0.940(.029) & 0.692(.104) & 0.957(.006) & 0.872(.010) \\ \cline{2-10}
\rule[-1ex]{0pt}{3.5ex} & Both (b) & 0.969(.002) & 0.804(.008) & 0.957(.004) & 0.876(.009) & 0.965(.011) & 0.783(.050) & 0.956(.002) & 0.877(.008) \\ \cline{2-10}
\rule[-1ex]{0pt}{3.5ex} & Real BCDR & 0.942(.033) & 0.713(.132) & \textbf{0.962(.003)} & \textbf{0.897(.001)} & 0.854(.001) & 0.500(.002) & 0.952(.002) & 0.873(.001) \\
\hline
\end{tabular}
}
\end{center}
\end{table}

\begin{table}[ht]
\caption{Results for classification of \textbf{tumour masses} using CNN (top) and Swin Transformer (bottom). 
The left-most column refers to the source of data added to the INbreast training set. 
``Both (a)'' means augmenting with synthetic patches from both BCDR (WGAN-GP) and OPTIMAM (DCGAN), while ``Both (b)'' means BCDR (DCGAN) and OPTIMAM (DCGAN). 
All classifiers are tested on the INbreast testset. The best results per column are presented in bold font. 
Results are shown as mean(std).} 
\label{tab:results2}
\begin{center}
\scalebox{0.70}{
\begin{tabular}{|c|l||c|c|c|c||c|c|c|c|} 
\hline
\rule[-1ex]{0pt}{3.5ex} & & \multicolumn{4}{c||}{100\% of the INbreast training data} & \multicolumn{4}{c|}{50\% of the INbreast training data} \\ \cline{2-10}
\rule[-1ex]{0pt}{3.5ex} & Augmentation & Accuracy & F1 & AUROC & AUPRC & Accuracy & F1 & AUROC & AUPRC  \\ \hline
\hline
\multirow{5}{*}{\parbox[t][12mm][b]{2.5mm}{\rotatebox[origin=c]{90}{CNN}}}
\rule[-1ex]{0pt}{3.5ex} & None & 0.895(.019) & 0.935(.015) & 0.962(.010) & 0.992(.002) & 0.897(.013) & 0.938(.009) & 0.955(.016) & 0.991(.003) \\ \cline{2-10}
\rule[-1ex]{0pt}{3.5ex} & BCDR (WGAN-GP) & \textbf{0.943(.023)} & \textbf{0.966(.014)} & 0.973(.010) & \textbf{0.995(.002)} & 0.872(.013) & 0.919(.008) & 0.962(.009) & 0.992(.002) \\ \cline{2-10}
\rule[-1ex]{0pt}{3.5ex} & BCDR (DCGAN) & 0.910(.016) & 0.944(.010) & \textbf{0.976(.005)} & \textbf{0.995(.001)} & 0.904(.020) & 0.941(.014) & \textbf{0.968(.014)} & \textbf{0.993(.003)} \\ \cline{2-10}
\rule[-1ex]{0pt}{3.5ex} & OPTIMAM (DCGAN) & 0.876(.072) & 0.919(.051) & 0.971(.010) & 0.994(.002) & 0.925(.033) & 0.954(.022) & \textbf{0.968(.008)} & \textbf{0.993(.002)} \\ \cline{2-10}
\rule[-1ex]{0pt}{3.5ex} & Both (a) & 0.908(.062) & 0.942(.043) & 0.970(.007) & 0.994(.002) & \textbf{0.933(.020)} & \textbf{0.960(.012)} & 0.946(.026) & 0.983(.011) \\ \cline{2-10}
\rule[-1ex]{0pt}{3.5ex} & Both (b) & 0.925(.019) & 0.954(.012) & 0.975(.001) & \textbf{0.995(.000)} & \textbf{0.933(.024)} & 0.959(.015) & 0.967(.009) & \textbf{0.993(.002)} \\ \cline{2-10}
\rule[-1ex]{0pt}{3.5ex} & Real BCDR & 0.912(.011) & 0.946(.006) & 0.968(.004) & 0.994(.001) & 0.885(.024) & 0.931(.014) & 0.935(.012) & 0.987(.002) \\
\hline
\hline
\multirow{5}{*}{\parbox[t][16mm][b]{2.5mm}{\rotatebox[origin=c]{90}{Swin Transformer}}}
\rule[-1ex]{0pt}{3.5ex} & None & 0.834(.019) & 0.891(.014) & 0.928(.015) & 0.986(.003) & 0.784(.010) & 0.860(.003) & 0.877(.015) & 0.976(.003) \\ \cline{2-10}
\rule[-1ex]{0pt}{3.5ex} & BCDR (WGAN-GP) & \textbf{0.950(.041)} & \textbf{0.969(.025)} & \textbf{0.978(.011)} & \textbf{0.996(.002)} & 0.933(.020) & 0.959(.013) & \textbf{0.973(.003)} & \textbf{0.995(.000)} \\ \cline{2-10}
\rule[-1ex]{0pt}{3.5ex} & BCDR (DCGAN) & 0.876(.053) & 0.920(.038) & 0.959(.029) & 0.992(.006) & 0.922(.010) & 0.953(.006) & 0.968(.006) & 0.994(.001) \\ \cline{2-10}
\rule[-1ex]{0pt}{3.5ex} & OPTIMAM (DCGAN) & 0.941(.020) & 0.964(.013) & 0.975(.008) & 0.995(.002) & 0.920(.018) & 0.952(.012) & 0.963(.013) & 0.993(.003) \\ \cline{2-10}
\rule[-1ex]{0pt}{3.5ex} & Both (a) & 0.933(.018) & 0.959(.012) & 0.973(.012) & 0.995(.002) & 0.933(.010) & 0.959(.007) & 0.972(.002) & 0.994(.000) \\ \cline{2-10}
\rule[-1ex]{0pt}{3.5ex} & Both (b) & 0.914(.032) & 0.947(.021) & 0.972(.016) & 0.995(.003) & \textbf{0.935(.010)} & \textbf{0.960(.006)} & 0.970(.001) & 0.994(.000) \\ \cline{2-10}
\rule[-1ex]{0pt}{3.5ex} & Real BCDR & 0.841(.010) & 0.907(.006) & 0.929(.007) & 0.987(.001) & 0.799(.006) & 0.862(.004) & 0.890(.006) & 0.979(.001) \\ \cline{2-10}
\hline
\end{tabular}
}
\end{center}
\end{table}

Most importantly, all experiments with synthetic patches exhibit an improved performance of the classifier compared to using only INbreast.
For example, when training a Swin Transformer on 50\% of available INbreast training data and all lesion types, the classifier reaches an F1-score of $0.734$.
However, when adding synthetic patches from a DCGAN trained on BCDR and the same number of healthy samples from INbreast, the classifier reaches an F1-score of $0.880$.
Therefore, providing the classifier with synthetic data from another dataset improves its F1-score by $0.146$ in this case.
This effect tends to be strongest in the low-data regime, i.e. 50\% of training data and only masses, and is particularly pronounced in the Swin Transformer.

\section{DISCUSSION}
\label{sec:discussion}

In this work, we tested the hypothesis that sharing generative models instead of private patient data across centres is a beneficial alternative. We showed this empirically by comparing classification model performance when trained on synthetic data sampled from the shared generative models. 

The results show that additional synthetic training data improves the performance of both classification models, where the improvement for the Swin Transformer is more pronounced than for the CNN. This is due to the larger number of trainable parameters in Swin Transformers (28M) 
compared to our CNN (1.1M). Additionally, since transformers do not possess inductive biases as CNNs, they require more training data to learn internal relations within the image.
Therefore, the Swin Transformer is strongly benefiting from additional training data provided by the generators. In line with this argument the performance increase is most notable in the low-data regime, namely, where the training data is reduced by 50\% and where only masses are used, as opposed to all lesion types.

Observing that adding the same number of synthetic patches from multiple sources, i.e. 50\% BCDR and 50\% OPTIMAM, does rarely improve upon adding 100\% of a single-source (either BCDR or OPTIMAM). We hypothesise that as we increase the variation with multi-sources while leaving the dataset size constant, the classifier might have to learn/overcome additional domain shifts \cite{garrucho2022domain} and variation. For instance, OPTIMAM mass patches contain less background than BCDR mass patches while also being based on different acquistion protocols.

Furthermore, we note that the results often worsen when adding the real BCDR patches as compared to adding synthetic (BCDR) patches for both all lesions and only masses experiments. We suppose part of the performance decay stems from a domain-shift between BCDR and INbreast that could translate less into the GAN-generated BCDR synthetic data. We leave further investigation into this aspect to future work and point out that Garrucho et al \cite{garrucho2022domain} describe a considerable distribution-shift between BCDR and other mammography datasets, such as INbreast, in image contrast and intensity, as well as in the distribution of lesions, in terms of their size and aspect ratio.


As our results show promising potential of utilising GANs for privacy-preserving data sharing strategies, we also note that our proposed approach depends heavily on the willingness of centres to train and share generative models of their private datasets. In this context, we highlight the need for additional privacy preserving measures and thorough investigation as to how the shared generative model can leak private patient information (i) in general and (ii) when subjected to training data reconstruction attacks. One counter-measure against leakage of private information that we leave for future work is applying differential privacy~\cite{10.1007/11787006_1} to GAN training, which provides a privacy guarantee for the GAN training data~\cite{Osuala2021}. Before sharing a model, it is to be assessed whether private attributes such as particular patient-identifying anatomical or pathological features can be extracted from the shared model's parameters or predictions. For instance, model inversion attacks have shown to successfully reveal identifying imaging features from the training data.

To further validate the benefits of inter-centre generative model sharing, we recommend future work to test prediction performance improvement on full mammograms apart from patches and on further datasets across (imaging and non-imaging) modalities, organs, and standardised clinical tasks \cite{lekadir2021future} with universal classification objectives (e.g., benign/malignant apart from healthy/non-healthy classification). We further propose future work to explore the effect of initialisation of classification models with pretrained weights alongside the effect of traditional preprocessing and data augmentation techniques, such as histogram and intensity scale normalization techniques~\cite{garrucho2022domain}. Apart from classification, validation on further downstream tasks such as object detection, semantic segmentation, and domain-adaptation can reveal further insights into the advantages and limitations of generative model sharing. Lastly, further recommendations can be derived from the evaluation and comparison of additional types of GANs and generative models in diverse settings with varying patient data resource constraints between centres.

\section*{ACKNOWLEDGMENTS}
This project has received funding from the European Union’s Horizon 2020 research and innovation programme under grant agreement No 952103.

\bibliographystyle{plain} 
\bibliography{references} 

\begin{thebibliography}{10}

\bibitem{Abdelrahman2021}
Leila Abdelrahman, Manal {Al Ghamdi}, Fernando Collado-Mesa, and Mohamed
  Abdel-Mottaleb.
\newblock {Convolutional neural networks for breast cancer detection in
  mammography: A survey}.
\newblock {\em Computers in Biology and Medicine}, 131(January):104248, 2021.

\bibitem{reluAgarap2018deep}
Abien~Fred Agarap.
\newblock Deep learning using rectified linear units (relu).
\newblock {\em arXiv preprint arXiv:1803.08375}, 2018.

\bibitem{alyafi20quality}
Basel Alyafi, Oliver Diaz, Premkumar Elangovan, Joan~C Vilanova, Javier del
  Riego, and Robert Marti.
\newblock Quality analysis of dcgan-generated mammography lesions.
\newblock In {\em 15th International Workshop on Breast Imaging (IWBI2020)},
  volume 11513, page 115130B. International Society for Optics and Photonics,
  2020.

\bibitem{alyafi2020dcgans}
Basel Alyafi, Oliver Diaz, and Robert Marti.
\newblock Dcgans for realistic breast mass augmentation in x-ray mammography.
\newblock In {\em Medical Imaging 2020: Computer-Aided Diagnosis}, volume
  11314, page 1131420. International Society for Optics and Photonics, 2020.

\bibitem{arjovsky2017wasserstein}
Martin Arjovsky, Soumith Chintala, and L{\'e}on Bottou.
\newblock Wasserstein generative adversarial networks.
\newblock In {\em International conference on machine learning}, pages
  214--223. PMLR, 2017.

\bibitem{becker2017deep}
Anton~S Becker, Magda Marcon, Soleen Ghafoor, Moritz~C Wurnig, Thomas
  Frauenfelder, and Andreas Boss.
\newblock Deep learning in mammography: diagnostic accuracy of a multipurpose
  image analysis software in the detection of breast cancer.
\newblock {\em Investigative radiology}, 52(7):434--440, 2017.

\bibitem{bi2019artificial}
Wenya~Linda Bi, Ahmed Hosny, Matthew~B Schabath, Maryellen~L Giger, Nicolai~J
  Birkbak, Alireza Mehrtash, Tavis Allison, Omar Arnaout, Christopher Abbosh,
  Ian~F Dunn, et~al.
\newblock Artificial intelligence in cancer imaging: clinical challenges and
  applications.
\newblock {\em CA: a cancer journal for clinicians}, 69(2):127--157, 2019.

\bibitem{castro2020causality}
Daniel~C Castro, Ian Walker, and Ben Glocker.
\newblock Causality matters in medical imaging.
\newblock {\em Nature Communications}, 11(1):1--10, 2020.

\bibitem{10.1007/11787006_1}
Cynthia Dwork.
\newblock Differential privacy.
\newblock In Michele Bugliesi, Bart Preneel, Vladimiro Sassone, and Ingo
  Wegener, editors, {\em Automata, Languages and Programming}, pages 1--12,
  Berlin, Heidelberg, 2006. Springer Berlin Heidelberg.

\bibitem{Fukushima1980}
Kunihiko Fukushima.
\newblock {Neocognitron: A self-organizing neural network model for a mechanism
  of pattern recognition unaffected by shift in position}.
\newblock {\em Biological Cybernetics}, 36(4):193--202, 1980.

\bibitem{garrucho2022domain}
Lidia Garrucho, Kaisar Kushibar, Socayna Jouide, Oliver Diaz, Laura Igual, and
  Karim Lekadir.
\newblock Domain generalization in deep learning-based mass detection in
  mammography: A large-scale multi-center study.
\newblock {\em arXiv preprint arXiv:2201.11620}, 2022.

\bibitem{globalCancerObservatory}
{Global Cancer Observatory}.
\newblock The global cancer observatory (gco) is an interactive web-based
  platform presenting global cancer statistics to inform cancer control and
  research.
\newblock \url{https://gco.iarc.fr/}, 2021.
\newblock Accessed: 2021-11-25.

\bibitem{goodfellow2014generative}
Ian Goodfellow, Jean Pouget-Abadie, Mehdi Mirza, Bing Xu, David Warde-Farley,
  Sherjil Ozair, Aaron Courville, and Yoshua Bengio.
\newblock Generative adversarial nets.
\newblock In {\em Advances in neural information processing systems}, pages
  2672--2680, 2014.

\bibitem{gulrajani2017improved}
Ishaan Gulrajani, Faruk Ahmed, Martin Arjovsky, Vincent Dumoulin, and Aaron
  Courville.
\newblock Improved training of wasserstein gans.
\newblock {\em arXiv preprint arXiv:1704.00028}, 2017.

\bibitem{halling2020optimam}
Mark~D Halling-Brown, Lucy~M Warren, Dominic Ward, Emma Lewis, Alistair
  Mackenzie, Matthew~G Wallis, Louise~S Wilkinson, Rosalind~M Given-Wilson,
  Rita McAvinchey, and Kenneth~C Young.
\newblock Optimam mammography image database: A large-scale resource of
  mammography images and clinical data.
\newblock {\em Radiology: Artificial Intelligence}, page e200103, 2020.

\bibitem{ioffe2015batch}
Sergey Ioffe and Christian Szegedy.
\newblock Batch normalization: Accelerating deep network training by reducing
  internal covariate shift.
\newblock In {\em International conference on machine learning}, pages
  448--456. PMLR, 2015.

\bibitem{lecun1998gradient}
Yann LeCun, L{\'e}on Bottou, Yoshua Bengio, and Patrick Haffner.
\newblock Gradient-based learning applied to document recognition.
\newblock {\em Proceedings of the IEEE}, 86(11):2278--2324, 1998.

\bibitem{lekadir2021future}
Karim Lekadir, Richard Osuala, Catherine Gallin, Noussair Lazrak, Kaisar
  Kushibar, Gianna Tsakou, Susanna Auss{\'o}, Leonor~Cerd{\'a} Alberich, Kostas
  Marias, Manolis Tsiknakis, et~al.
\newblock Future-ai: Guiding principles and consensus recommendations for
  trustworthy artificial intelligence in medical imaging.
\newblock {\em arXiv preprint arXiv:2109.09658}, 2021.

\bibitem{liu2021swin}
Ze~Liu, Yutong Lin, Yue Cao, Han Hu, Yixuan Wei, Zheng Zhang, Stephen Lin, and
  Baining Guo.
\newblock Swin transformer: Hierarchical vision transformer using shifted
  windows.
\newblock {\em arXiv preprint arXiv:2103.14030}, 2021.

\bibitem{Lopez2012}
Miguel~G Lopez, Naimy Posada, Daniel~C Moura, Ra{\'u}l~Ramos Poll{\'a}n,
  Jos{\'e} M~Franco Valiente, C{\'e}sar~Su{\'a}rez Ortega, Manuel Solar,
  Guillermo Diaz-Herrero, IMAP Ramos, Joana Loureiro, et~al.
\newblock Bcdr: a breast cancer digital repository.
\newblock In {\em 15th International conference on experimental mechanics},
  volume 1215, 2012.

\bibitem{magister2021generative}
Lucie~Charlotte Magister and Ognjen Arandjelovi{\'c}.
\newblock Generative image inpainting for retinal images using generative
  adversarial networks.
\newblock In {\em 2021 43rd Annual International Conference of the IEEE
  Engineering in Medicine \& Biology Society (EMBC)}, pages 2835--2838. IEEE,
  2021.

\bibitem{Moreira2012}
In{\^e}s~C Moreira, Igor Amaral, In{\^e}s Domingues, Ant{\'o}nio Cardoso,
  Maria~Joao Cardoso, and Jaime~S Cardoso.
\newblock Inbreast: toward a full-field digital mammographic database.
\newblock {\em Academic radiology}, 19(2):236--248, 2012.

\bibitem{Osuala2021}
Richard Osuala, Kaisar Kushibar, Lidia Garrucho, Akis Linardos, Zuzanna
  Szafranowska, Stefan Klein, Ben Glocker, Oliver Diaz, and Karim Lekadir.
\newblock A review of generative adversarial networks in cancer imaging: New
  applications, new solutions.
\newblock {\em arXiv preprint arXiv:2107.09543}, 2021.

\bibitem{richard_osuala_2022_6327625}
Richard Osuala, Noussair Lazrak, Kaisar Kushibar, Lidia Garucho, Socayna
  Jouide, Grzegorz Skorupko, Oliver Diaz, and Karim Lekadir.
\newblock {medigan: Synthetic Medical Data From Pretrained Generative Models},
  March 2022.

\bibitem{pytorchPaszke}
Adam Paszke, Sam Gross, Francisco Massa, Adam Lerer, James Bradbury, Gregory
  Chanan, Trevor Killeen, Zeming Lin, Natalia Gimelshein, Luca Antiga, Alban
  Desmaison, Andreas Kopf, Edward Yang, Zachary DeVito, Martin Raison, Alykhan
  Tejani, Sasank Chilamkurthy, Benoit Steiner, Lu~Fang, Junjie Bai, and Soumith
  Chintala.
\newblock Pytorch: An imperative style, high-performance deep learning library.
\newblock In H.~Wallach, H.~Larochelle, A.~Beygelzimer, F.~d\textquotesingle
  Alch\'{e}-Buc, E.~Fox, and R.~Garnett, editors, {\em Advances in Neural
  Information Processing Systems 32}, pages 8024--8035. Curran Associates,
  Inc., 2019.

\bibitem{radford2015unsupervised}
Alec Radford, Luke Metz, and Soumith Chintala.
\newblock Unsupervised representation learning with deep convolutional
  generative adversarial networks.
\newblock {\em arXiv preprint arXiv:1511.06434}, 2015.

\bibitem{shin2018medical}
Hoo-Chang Shin, Neil~A Tenenholtz, Jameson~K Rogers, Christopher~G Schwarz,
  Matthew~L Senjem, Jeffrey~L Gunter, Katherine~P Andriole, and Mark Michalski.
\newblock Medical image synthesis for data augmentation and anonymization using
  generative adversarial networks.
\newblock In {\em International workshop on simulation and synthesis in medical
  imaging}, pages 1--11. Springer, 2018.

\bibitem{dropout}
Nitish Srivastava, Geoffrey Hinton, Alex Krizhevsky, Ilya Sutskever, and Ruslan
  Salakhutdinov.
\newblock Dropout: A simple way to prevent neural networks from overfitting.
\newblock {\em Journal of Machine Learning Research}, 15:1929--1958, 06 2014.

\bibitem{vaswani2017attention}
Ashish Vaswani, Noam Shazeer, Niki Parmar, Jakob Uszkoreit, Llion Jones,
  Aidan~N Gomez, Lukasz Kaiser, and Illia Polosukhin.
\newblock {Attention is all you need}.
\newblock {\em arXiv preprint arXiv:1706.03762}, 2017.

\end{thebibliography}

\end{document}